\documentclass[aps,prd,superscriptaddress,showpacs,preprintnumbers]{revtex4}
\usepackage{graphicx}

\begin{document}

\parskip=0.3cm

%\begin{titlepage}

\title{Dual-Regge Approach to High-Energy,
Low-Mass Diffraction Dissociation}

\author{L.L. Jenkovszky}
\affiliation{Bogolyubov Institute for Theoretical Physics (BITP),
Ukrainian National Academy of Sciences \\14-b, Metrolohichna str.,
Kiev, 03680, Ukraine}

\author{O.E. Kuprash}
\affiliation{Taras Shevchenko National University, Kiev, Ukraine}

\author{J.W.~L\"ams\"a}
\affiliation{Physics Department, Iowa State University,
Ames, 50011 Iowa, USA}

\author{V.K.~Magas}
\affiliation{Departament d'Estructura i Constituents de la
Mat\'eria,\\ Universitat de Barcelona, Diagonal 647, 08028 Barcelona, Spain}

\author{R.~Orava}
\affiliation{Helsinki Institute of Physics, Division of Elementary Particle Physics,\\
P.O. Box 64 (Gustaf H\"allstr\"ominkatu 2a), FI-00014 University of Helsinki, Finland}
\affiliation{CERN, CH-1211 Geneva 23, Switzerland}

\begin{abstract}
A dual-Regge model with a nonlinear proton Regge trajectory in the
missing mass $(M^2_X)$ channel, describing the experimental data
on low-mass single diffraction dissociation (SDD), is constructed.
Predictions for the LHC energies are given.
\end{abstract}

\pacs{11.55.-m, 11.55.Jy, 12.40.Nn}

\maketitle

%\end{titlepage}
\section{Introduction} \label{s1}
The possible existence of a new class of processes, later named
diffraction dissociation, for the first time was indicated in
1953, in a short paper by Pomeranchuk and Feinberg \cite{P-F}. The
possibility of observing diffractive inelastic processes producing
states $X$ of large mass was studied subsequently, in 1960 by Good
and Walker \cite{G-W} (for a review see Ref. \cite{Zotov}).

Experimentally, diffraction dissociation in proton-proton
scattering was intensively studied in the '70-ies at the Fermilab
and the CERN ISR \cite{Scham1, Albrow, Scham2}. In particular, in
Ref. \cite{Scham2} double differential cross section
${d\sigma}\over{dtdM_X^2}$ was measured in the region
$0.024<-t<0.234$ (GeV/c)$^2,\ \ 0<M^2<0.12s$, and $(105<s<752)$
GeV$^2$, and a single peak in $M_X^2$ was identified.

Low-mass diffraction dissociation (DD) of protons, single
\begin{equation}\label{SD}
    pp\rightarrow pX,
\end{equation}
and double, are among the priorities at the LHC.

For the CMS Collaboration, the SDD mass coverage is presently
limited to some 10 GeV. With the Zero Degree Calorimeter (ZDS),
this could be reduced to smaller masses, in case the SDD system
produces very forward neutrals, i.e. like a $N^*$ decaying into a
fast leading neutron. Together with the T2 detectors of TOTEM, SDD
masses down to 4 GeV could be covered. This is not the case until
TOTEM trigger (data acquisition) system are combined together with
the CMS ones. This is not likely before the year 2012 shut down.
In principle ATLAS can do similar improvement, since the LHC
lay-out at the distance of our proposed Forward Shower Counters'
(FSC) locations is similar. ALICE and LHCb have different beam
arrangements, but their acceptances for central diffraction
(double pomeron exchange) was also investigated, see, e.g.,
\cite{Risto}).

While high-mass diffraction dissociation (DD) receives much
attention, mainly due to its relatively easy theoretical treatment
within the triple Reggeon formalism \cite{Goulianos1, Roy, JL, DL}
and successful reproduction  of the data \cite{Goulianos1,
Montanha}, this is not the case for low-masses, which are beyond
the range of perturbative quantum chromodynamics (QCD).  The
forthcoming measurements at the LHC urge a relevant theoretical
understanding and treatment of low mass DD, which essentially has
both spectroscopic and dynamic aspects. The low-mass, $M_X$
spectrum is rich of nucleon resonances. Their discrimination is a
difficult experimental task, and theoretical predictions of the
appearance of the resonances depending on $s,\ t$ and $M$ is also
very difficult since, as mentioned, perturbative QCD, or
asymptotic Regge pole formula are of no use here. With this paper
we try to partially fill this gap, attacking the problem by means
of a dual-Regge approach to the inelastic form factor (production
amplitude) in which non-linear Regge trajectories play an
essential role.

We start with single diffraction dissociation (SDD).
Generalization to double diffraction dissociation (DDD) is
straightforward.

\begin{figure}[htb]
\includegraphics[width=.75\textwidth]{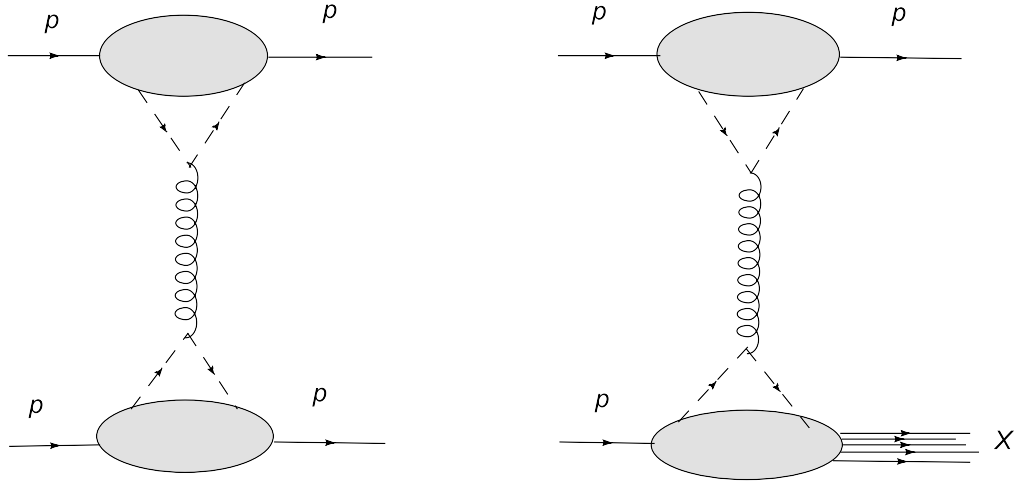}
\caption{Figure caption: Elastic scattering (left panel)
and diffraction dissociation (right panel) in a model with a
pomeron exchange coupled to the proton by quarks.}
\label{fig:diagram}
\end{figure}

\begin{figure}[htb]
 \includegraphics[width=.4\textwidth]{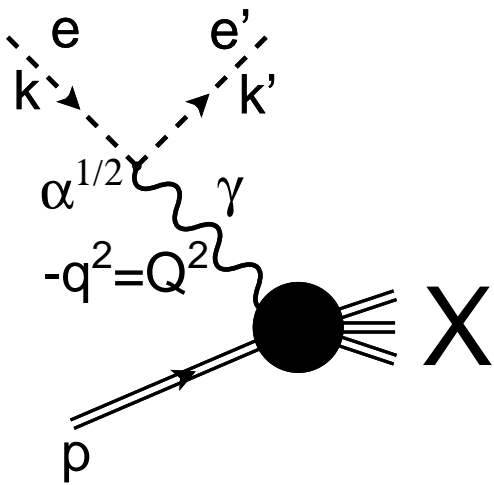}
\caption{Virtual photon+proton $\rightarrow M_X$
transition.}
\label{fig:drow1}
\end{figure}

Diffraction, elastic and inelastic, in the LHC energy range is
dominated by a single Pomeron exchange in the $t$ channel (see
e.g. \cite{L, FJOPPS}), enabling the use of Regge factorization, Fig.
\ref{fig:diagram}. Accordingly, the knowledge of two vertices and
the Regge propagator is essential for the construction of the
scattering amplitude. Relying on the known properties of the
elastic proton-Pomeron-proton vertex and by adopting a simple
supercritical Pomeron pole exchange (propagator) in the $t$
channel, we concentrate on the construction of a proper inelastic
proton-Pomeron-$M_X$ vertex, the central object of our study. The
solution of this problem, to large extent, became possible due to
the similarity between the inelastic $\gamma^* p\rightarrow M_x$
and Pomeron+proton$\rightarrow M_x$ vertices. We will extensively
use the earlier results on the $\gamma^* p\rightarrow M_x$
transition, successfully applied to the JLab data \cite{Magas,
complete} in constructing the lower, Pomeron+proton$\rightarrow
M_x$ vertices of Fig. \ref{fig:diagram}, right panel. In doing
so, we draw a parallel between the virtual photon and the Pomeron.
They are similar, $W_2(q^2,s)_{\gamma^*p\rightarrow N_i^*,\Delta}\
{\rm (at\ JLab)} \rightarrow W_2(t,M^2)_{Pp\rightarrow N^*}\ ({\rm
at\ the\ LHC)}$, apart from their opposite $C$ parities, and, of
course, one should remember about the changes in kinematics: the
photon virtuality (e.g., at the JLab) $Q^2$, etc of
Fig.\ref{fig:drow1} here becomes the squared momentum transfer $t$
of Fig. \ref{fig:diagram}.

The unknown inelastic form factor of the type shown in Fig.
\ref{fig:drow1}, by the optical theorem, is related to the
imaginary part of the forward $\gamma^*(P)-p$ scattering
amplitude. Following Refs. \cite{Predazzi, Magas, complete}, we
use a dual amplitude for this reaction, in its low-energy (here:
missing mass), resonance region, dominated by the contribution of
relevant direct-channel trajectories. The correct choice of these
trajectories is a crucial point in our approach. In the case of
$\gamma^* p$ scattering (e.g., JLab) these were the $N^*$ and
$\Delta$ trajectories, see Refs. \cite{Predazzi, Magas, complete}.
Here, instead, by quantum numbers, the relevant direct channel
trajectory is that of the proton, to be parametrized in Sec.
\ref{Sec:trajectory}.

    In principle, one could proceed by counting the resonances
one-by-one; however, apart from the technical complexity of
counting single resonances, there is also a conceptual one: Regge
trajectories and, more generally, dual models comprise the
dynamics in a complete and continuous way, thus opening the way to
study and relate different reactions in any kinematical region.
Examples are finite mass sum rules, contained in the present
formalism automatically. One more important point: the advantage
of using the dual-Regge model with non-linear Regge trajectory
presented in this paper over a one-to-one account for the
particular resonances is that it automatically takes care of the
relative weight of each resonance, and extrapolates to higher
masses, with a limited number of resonances on any trajectory.

\section{Elastic Scattering} \label{s2}
\vspace{-0.5cm}

The $pp$ scattering amplitude corresponding to Fig. \ref{fig:diagram} (left) is \cite{JL}
\begin{equation} \label{amplitude}
A(s,t)=-\beta^2[f^u(t)+f^d(t)]^2(s/s_0)^{\alpha_P(t)-1}\frac{1+e^{-i\pi\alpha_P(t)}}{\sin\pi\alpha_P(t)},
\end{equation}
where $f^u(t)$ and $f^d(t)$ are the amplitudes for the emission of
$u$ and $d$ valence quarks by the nucleon, $\beta$ is the
quark-Pomeron coupling, to be determined below; $\alpha_P(t)$ is a
vacuum Regge trajectory. It is assumed \cite{JL} that the Pomeron
couples to the proton via quarks like a scalar photon. Thus, the
unpolarized elastic $pp$ differential cross section is
\begin{equation}\label{sigma}
    \frac{d\sigma}{dt}=\frac{[3\beta F^p(t)]^4}{4\pi\sin^2[\pi\alpha_P(t)/2]}(s/s_0)^{2\alpha_P(t)-2}.
\end{equation}
The norm $\beta$ appearing in Eq. (\ref{amplitude}) was found in Ref.
\cite{JL} from the forward elastic scattering, $d\sigma/dt\approx
80$ mb/GeV$^2$ at $\sqrt{s}=23.6$ and $30.8$ GeV, resulting, at
unite Pomeron intercept, $\alpha_P(0)=1,$ in
$\beta^4/(4\pi)\approx 1$ mb/GeV$^2$ \cite{JL}.

To account for the rise of the cross sections, following the model
and fits of Donnachie and Landshoff, see \cite{L} and earlier
references therein, we use a Pomeron trajectory whose
intercept is slightly beyond one, namely, $\alpha_P(0)=1.08$
providing for excellent fits to the total cross sections \cite{L}.
However, the extrapolation with such an intercept and input value of $\beta$
strongly overshoots the elastic forward cross section measured at
higher energies, e.g. $\sqrt s=1800$ GeV \cite{Amos}. There are
several reasons for this inconsistency. One is that, at the
normalization point, $23.6$ or $30.6$ GeV, the contribution from
secondary Reggeons, and/or a constant background should be included.
In what follows we use the Pomeron trajectory of the form, see \cite{L}
$\alpha_P(t)=1.08+0.25t$ and consequently relax the above norm of $\beta$. Instead,
it will be included in the overall normalization factor
of the amplitude/cross section $A_0$, that
absorbs also the parameter $a$ of  eq. (\ref{ImA}), from Section \ref{Results}.

Another important issue is the neglect of absorption (unitary) corrections.
We intend to come back in a forthcoming investigation to the study of the role of the subleading reggeons
and of the absorption corrections .

A dipole form can be used for the form factor
\begin{equation}\label{FF1}
F^p(t)=\frac{4m^2-2.9t}{4m^2-t}\frac{1}{(1-t/0.71)^2},
\end{equation}
where $m$ is the proton mass.

\section{Single Diffraction Dissociation (SDD)}
\label{SDD}
In single diffraction dissociation, Eq. (\ref{SD}),
a system $X$ with a missing mass $M_X$ is produced at small $|t|$.
At sufficiently large $s/M_X^2$, which is the case at the LHC, the
process is dominated by a Pomeron exchange. This case was treated
in Ref. \cite{JL} for missing masses beyond the resonance region,
and in Ref. \cite{Ravndal} in the resonance region. For large
missing masses, the triple Regge limit applies \cite{Roy,
Goulianos, Montanha, Collins}. Although large-$M_X$ diffraction
dissociation is outside the scope of the present paper, we mention
it below, in particular in connection with duality relations
called finite mass sum rules, that relate low- and high missing
mass dynamics.

Similar to the case of elastic scattering (Sec. \ref{s2}), the
double differential cross section for the reaction (\ref{SD}), by
Regge factorization, can be written as
\begin{equation}
\label{DD1} \frac{d^2\sigma}{dtdM_X^2}\sim
\frac{9\beta^4[F^p(t)]^2}{4\pi\sin^2[\pi\alpha_P(t)/2]}(s/M_X^2)^{2\alpha_P(t)-2}
\Bigl[\frac{W_2}{2m}\Bigl(1-M_X^2/s\Bigr)-mW_1(t+2m^2)/s^2\Bigr],
\end{equation}
where $W_i,\ \ i=1,2$ are related to the structure functions of
the nucleon and $W_2\gg W_1.$ For high $M_X^2$, the $W_{1,2}$ are
Regge-behaved, while for small $M_X^2$ their behavior is dominated
by nucleon resonances. Thus, the behavior of (\ref{DD1}) in the
low missing mass region to a large extent depends on the
transition form factors or resonance structure functions. The
knowledge of the inelastic form factors (or transition amplitudes)
is crucial for the calculation of low-mass diffraction
dissociation from Eq. (\ref{DD1}). We introduce these transition
amplitudes  in the next section.

At large $s$ (the LHC energies), one can safely neglect terms
$M_X^2/s$ and $(t+2m^2)/s$ in Eq. (\ref{DD1}). Furthermore, we
have replaced the familiar form of the signature factor in the
amplitude, ${{1+e^{-i\pi\alpha_P(t)}}}\over{\sin\pi\alpha_P(t)}$,
used in \cite{JL}, by a simple exponential one
$e^{-i\pi\alpha_P(t)/2}.$ For the proton elastic form factor
$F^p(t)$, eq. (\ref{FF1}), we use a dipole form
\begin{equation}
\label{F_t} F^p(t)=(1-t/0.71)^{-2}\,.
\end{equation}
(note that here we neglect the first factor of Eq. (4) producing a
break in the small $|t|$ behavior of the elastic differential
cross section).

 Hence Eq.
(\ref{DD1}), in the LHC energy region simplifies to:
\begin{equation}
\label{DD2} \frac{d^2\sigma}{dtdM_X^2}\approx
\frac{9\beta^4[F^p(t)]^2}{4\pi}(s/M_X^2)^{2\alpha_P(t)-2}
\frac{W_2}{2m}\,.
\end{equation}

Eqs. (\ref{DD1}) and (\ref{DD2}) do not contain the elastic
scattering limit because the inelastic form factor $W_2(M_X,t) $
has no elastic form factor limit $F(t)$ as $M_X\rightarrow m$.
This problem is similar to the $x\rightarrow 1$ limit of the deep
inelastic structure function $F_2(x,Q^2).$ The elastic
contribution to SDD should be added separately, as discussed
below, in Sec. \ref{Results}. To be sure, we eliminate in the present
work this region by imposing $M_X^2>2$ GeV$^2$.

\section{Transition Form Factors}
\label{sec_trans}

The  one-by-one account for single resonances is a possible,
although not efficient for the calculation of the SD cross
section, to which, at low missing masses, a sequence of many
resonances contribute. The definition and identification of these
resonances is not unique; moreover with increasing masses (still
within "low-mass diffraction"), they gradually disappear. Similar
to the case of electroproduction, the (dis)appearance of
resonances in the cross section depends on two variables, their
mass and the virtuality or the "probe" (photon with $Q^2$ in
electroproduction and Pomeron with $t$ in SDD). The finite widths
of the resonances can be introduced by a replacement
\cite{Carlson}:
$$\delta(W^2-m_N^2)\approx
\frac{1}{2m_R\pi}\frac{\Gamma_R/2}{(W-m_R)^2+\Gamma^2_R/4}\,,
$$
 where $\Gamma_R$ is the widths of the
resonance. At a resonance, $W=m_R$, the peak goes as high as
$1/(\pi m_R\Gamma_R)$.

A way to account for many resonances was suggested in paper
\cite{Magas}, based on the ideas of duality with a limited number
of resonances lying on non-linear Regge trajectories. This
approach was used \cite{complete} in a kinematically complete
analyses of the CLAS data from the JLab on the proton structure
function. The similarity between electroproduction of resonances
(e.g. at JLab) and low-mass SDD is the key point of our model. The
inelastic form factor (transition amplitude), the main ingradient
of the model, is constructed by analogy with the nucleon
resonances electroproduction amplitude. In both cases many
resonances overlap and their appearance depends both on the
reaction energy, which is replayced here by missing mass ($s\Rightarrow M_X^2$), and 
virtuality of the incident probe,  which is replayced here by the Pomoron's momentum transfer ($Q^2\Rightarrow -t$). This
interplay makes the problem complicated and interesting.

\subsection{Dual amplitude with Mandelstam
analyticity}\label{sec_DAMA}

\begin{figure}[htb]
 \includegraphics[width=.9\textwidth]{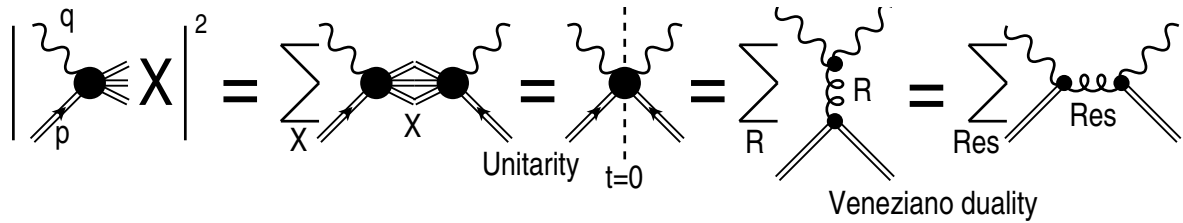}
\caption{Connection, through unitarity (generalized optical
theorem) and Veneziano-duality, between the inelastic form factor
and the sum of direct-channel resonances.} \label{fig:Regge-dual}
\end{figure}

The main idea behind the present work is the Regge-dual connection
between the inelastic form factor, Fig. \ref{fig:drow1}, appearing
in the lower vertex of Fig. \ref{fig:diagram}, and the
direct-channel, low energy (here: missing mass) dual amplitude, as
illustrated in Fig. \ref{fig:Regge-dual}.

Fig. \ref{fig:Regge-dual} shows the connection between the
inelastic form factor (structure function) appearing in the lower
vertex of the left panel of Fig. \ref{fig:diagram}, via duality,
unitarity (generalized optical theorem) and Veneziano-duality, and
its direct channel, resonance decomposition (rightmost term of
Fig. \ref{fig:Regge-dual}).

The invariant  on-shell scattering dual amplitude with Mandelstam
analyticity (DAMA), applicable  both to the diffractive and
non-diffractive components, reads \cite{DAMA, Predazzi, Magas,
complete}
\begin{equation}\label{D} D(s,t)=\int_0^1 {dz \biggl({z \over g}
\biggr)^{-\alpha_s(s')-1} \biggl({1-z \over
g}\biggr)^{-\alpha_t(t')-1}}, \end{equation} where $s'=s(1-z), \ \
t'=tz, \ \ g$ is a parameter, $g>1$, and $s, \ \ t$ are the
Mandelstam variables.

For $s\rightarrow\infty$ and fixed $t$ it has the following Regge
asymptotic behaviour \begin{equation}\label{Regge}
D(s,t)\approx\sqrt{{2\pi\over{\alpha_t(0)}}}g^{1+a+ib}\Biggl({s\alpha'_s(0)g\ln
g\over{\alpha_t(0)}}\Biggl)^{\alpha_t(0)-1}, \end{equation} where
$a=Re\ \alpha\Bigl({\alpha_t(0)\over{\alpha'_s(0)\ln g}}\Bigr)$ and
$b=Im\ \alpha\Bigl({\alpha_t(0) \over{\alpha'_s(0)\ln g}}\Bigr)$.

Contrary to the Veneziano model, DAMA \cite{DAMA} not only allows
for, but rather requires the use of nonlinear complex trajectories
providing the resonance widths via the imaginary part of the
trajectory, and, in a special case of restricted real part of the
trajectory, resulting in a finite number of resonances. More
specifically, the asymptotic rise of the trajectories in DAMA is
limited by the important upper
bound $$ |{\alpha_s(s)\over{\sqrt s\ln s}}|\leq const, \ \
s\rightarrow\infty. $$

The pole structure of DAMA is similar to that of the Veneziano
model, except that multiple poles appear on daughter levels
\cite{DAMA, Predazzi, Magas, complete},
\begin{equation}
 D(s,t)=\sum_{n=0}^{\infty}
g^{n+1}\sum_{l=0}^{n}\frac{[-s\alpha'_s(s)]^{l}C_{n-l}(t)}
{[n-\alpha_s(s)]^{l+1}},\,.
\label{series}
\end{equation}
where $C_n(t)$ is the
residue, whose form is fixed by the $t$-channel Regge trajectory
(see \cite{DAMA})
\begin{equation}
C_l(t)=\frac{1}{l!}\frac{d^l}{dz^l}\left[\biggl({1-z \over
g}\biggr)^{-\alpha_t(tz)}\right]_{z=0}\,.
\label{p5}
\end{equation}
The presence of the multipoles, Eq. (\ref{series}), does not
contradict the theoretical postulates.  On the other hand, they
can be removed without any harm to the dual model by means the
so-called Van der Corput neutralizer  \cite{DAMA}, resulting in a
"Veneziano-like" pole structure:
\begin{equation}
D(s,t)= \sum_{n=0}^{\infty} {C_n \over{n-\alpha_s(s)}}\,.
\label{eq23}
\end{equation}

We disregard the symmetry (spin and isospin) properties of
the problem, concentrating on its dynamics.

The main problem is how to introduce $Q^2$-dependence
 in the dual model,
matching its Regge asymptotic behaviour and pole structure to
standard forms known from the literature. (This is the famous
problem of the off-mass-shall continuation of the $S$ matrix.)
Note that any correct identification of this $Q^2$-dependence in a
single asymptotic limit of the dual amplitude, by duality, will
extend it to other kinematical regions. In Refs. \cite{Predazzi,
Magas, complete} a solution combining Regge behaviour and Bjorken
scaling limits of the structure functions (or $Q^2$-dependent
$\gamma^*p$ cross sections) was suggested (for an alternative
solution see Ref. \cite{MDAMA}).

Let ua remind that below $Q^2$ (photon virtuality in
 electroproduction) will be replaced by $-t$ (Pomeron
 "virtuality"), and $s$ will be replaced by 
$M^2_X$  (the direct, $Pp$ channel "energy").

\subsection{Dual-Regge model of the inelastic form factors
(transition amplitudes)}
\label{sec_Dual_Reg}

For our purposes, i.e. for low-mass SDD, the direct-channel pole
decomposition of the dual amplitude (\ref{eq23}) is relevant.
Anticipating its application in SDD, we write it as \footnote
{Note that resonances on the proton trajectory appear with spins
$J=1/2,\ 5/2,\ 9/2,\ 13/2...$.}
\begin{equation}\label{Magas}
A(M_X^2,t)=a\sum_{n=0,1,...}\frac{f(t)^{2(n+1)}}{2n+0.5-\alpha(M_X^2)},
\end{equation}
where $\alpha(M_X^2)$ is a non-linear Regge trajectory in the
Pomeron-proton system,  $t$ is the squared transfer momentum in
the $Pp\rightarrow Pp$ reaction, and $a$ is the normalization
factor, which will be absorbed 
together with $\beta$ in the overall normalization coefficient $A_0$ to be fitted to the data, see Sec. \ref{Results}.

The form factor $f(t)$ appearing in the $Pp\rightarrow Pp$ system
should not be confused with $F^p(t)$ in Eq. (\ref{SD}) (the $ppP$
vertex). It is fixed by the dual model \cite{Predazzi, Magas,
complete, MDAMA}, in particular by the compatibility of its Regge
asymptotics with Bjorken scaling \cite{Predazzi, Magas, complete}
and reads
\begin{equation}
\label{ft2} f(t)=(1-t/t_0)^{-2},
\end{equation}
where $t_0$ is a parameter to be fitted to the data, for example,
by comparing the hight of the resonance peaks for different $t$.
However, since for the moment we have no data on differential SDD
cross section, for simplicity we set $t_0=0.71$ GeV$^2$, as in the
proton elastic elastic form factor, eq. (\ref{FF1}).

Notice that in Eq. (\ref{Magas}) this form factor enters with a
power $2(n+1)$ strongly damping higher spin resonances
contributions \footnote{In an alternative approach of  Ref.
\cite{MDAMA} form factors enter with the same power for all the
resonances on a given trajectory. The advantage of the models with
increasing powers of the form factors is that the poorly known
high spin resonance are strongly suppressed and thus do not affect
the final results.}.

The inelastic form form factor in diffraction dissociation is
similar to that in $\gamma^* p$, treated in ref. \cite{complete},
up to the replacement of the photon by a Pomeron, whose parity is
different from that of the photon. As a consequence, we have a
single direct channel resonance trajectory, that of the proton,
plus the exotic, nonresonance trajectory providing the background,
dual to the Pomeron exchange in the cross channel. The proton
trajectory was studied in details in ref. \cite{Paccanoni} and
will be introduced in the next section.

Then we proceed, see for example \cite{MDAMA}:
\begin{equation}
\label{eq10}
\nu W_2(M_X^2,t)=F_2(x,t)= \frac{4(-t)(1-x)^2}{\alpha\, (M_x^2-m^2)(1+4m^2x^2/(-t))^{3/2} }Im\, A(M_X^2,t)\,,
\end{equation}
where $\alpha$ is a fine structure constant, $\nu$ is defined via $2m\nu=M_x^2-m^2-t$, and $x=\frac{-t}{2m\nu}$ is a Bjorken variable.
Thus finally we have:
\begin{equation}
\label{eq11}
\frac{W_2(M_X^2,t)}{2m}= \frac{4x(1-x)^2}{\alpha\, (M_x^2-m^2)(1+4m^2x^2/(-t))^{3/2} }Im\, A(M_X^2,t)\,,
\end{equation}
The imaginary part of the transition
amplitude reads
\begin{equation} \label{ImA}
Im\, A(M_X^2,t)=a \sum_{n=0,1,...}\frac{[f(t)]^{2(n+1)} Im\,
\alpha(M^2_x)}{(2n+0.5-Re\, \alpha(M_X^2))^2+ (Im\,
\alpha(M_X^2))^2}.
\end{equation}

Next we insert the proton trajectory $\alpha(M_X^2)$ into Eq.
(\ref{ImA}), and subsequently into Eq. (\ref{DD2}). The explicit
expression for the proton trajectory and the values of the
parameters are presented in the next section. For more details see
also Ref. \cite{Paccanoni}.

\section{The proton trajectory in the $M^2_X$-channel}
\label{Sec:trajectory}

\begin{figure}[htb]
\includegraphics[width=.49\textwidth,bb= 10 140 540 660]{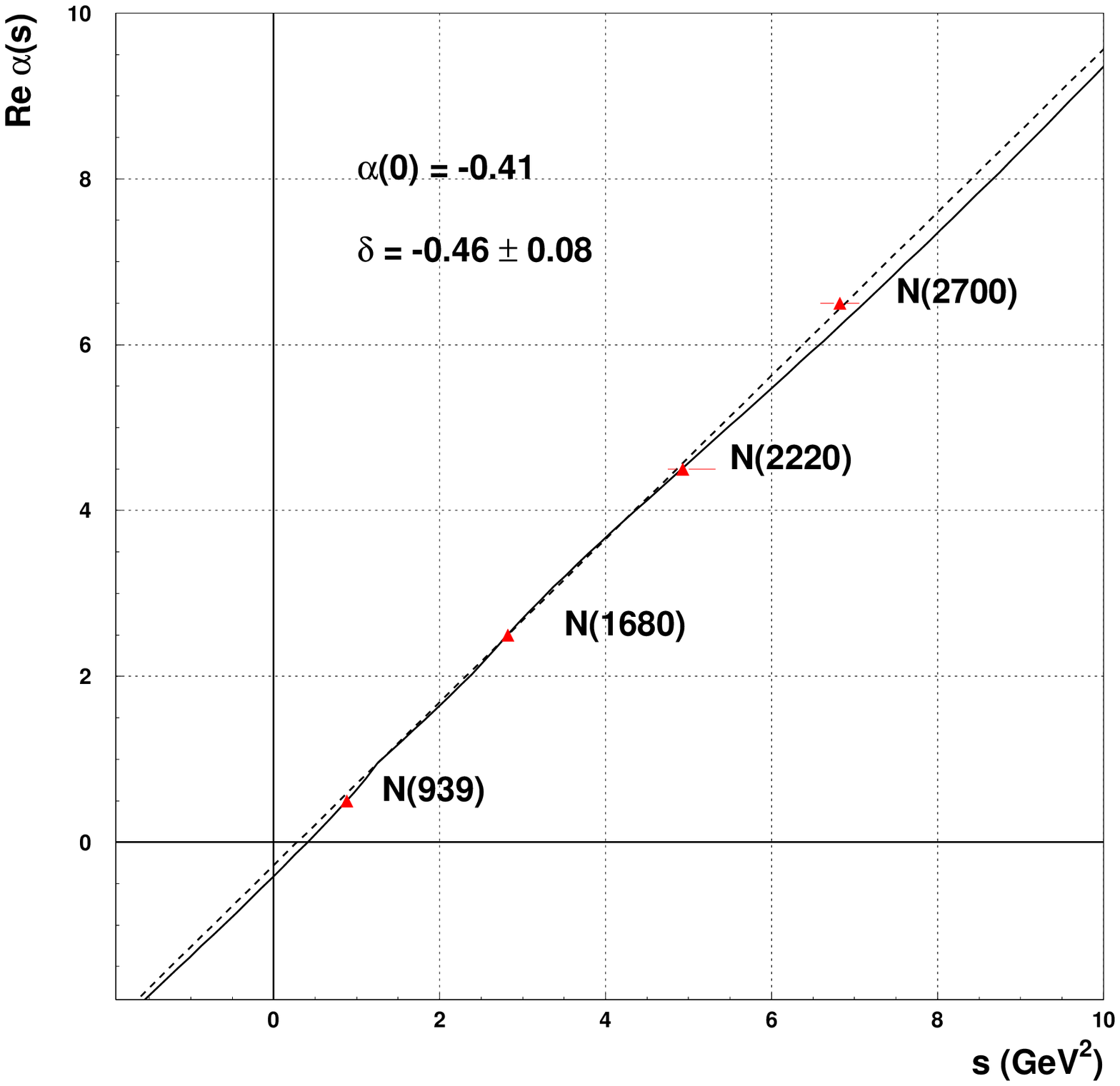}
\includegraphics[width=.49\textwidth,bb= 10 140 540 660]{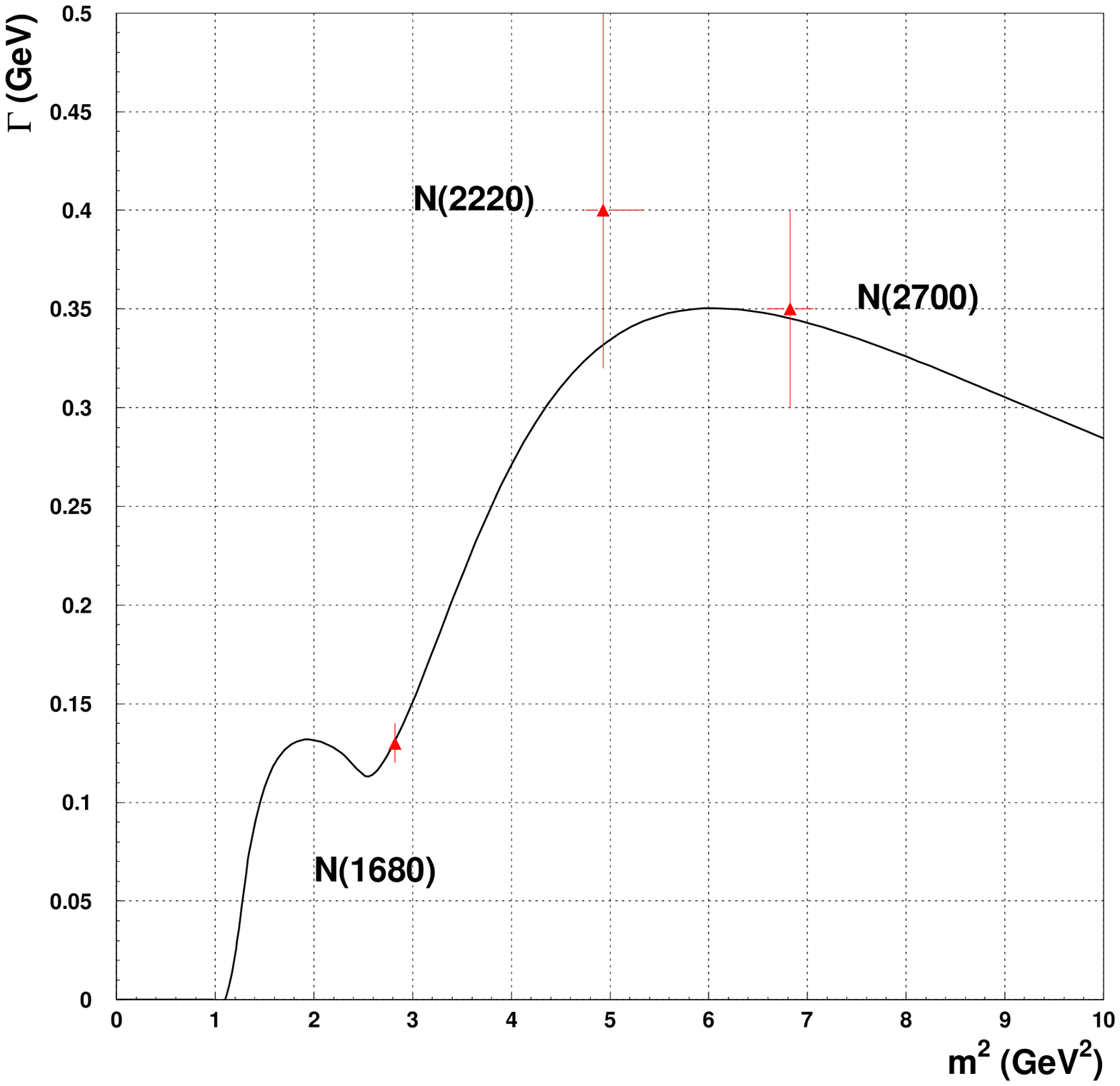}
\parbox[t]{8.5cm}{\caption{The real part of the proton Regge trajectory. The dashed line
corresponds to the result of a linear fit, the solid line is the
fit from \cite{Paccanoni}. \label{fig:n1}}}
\hfill~\parbox[t]{8.cm}{\caption{ The widths of the resonances,
$\Gamma=\frac{{\cal I}m\,\alpha(M^2)}{M{\cal R}e\,\alpha'(M^2)}~,$
appearing on the proton trajectory, calculated and fitted in
\cite{Paccanoni}. \label{fig:n1width}}}
\end{figure}

The Pomeron-proton channel, $Pp\rightarrow M_X^2$ (see the lower
part of Fig. \ref{fig:diagram}, right pannel) couples to the
proton trajectory, with the $I(J^P)$ resonances: $1/2(5/2^+),\
F_{15},\ m=1680$ MeV, $\Gamma=130$ MeV; $1/2(9/2^+),\ H_{19},\
m=2200$ MeV, $\Gamma=400$ MeV; and $1/2(13/2^+),\ K_{1,13},\
m=2700$ MeV, $\Gamma=350$ MeV. The status of the first two is
firmly established \cite{particles}, while the third one,
$N^*(2700),$ is less certain, with its width varying between
$350\pm 50$ and $900\pm 150$ MeV \cite{particles}. Still, with the
stable proton included, we have a fairly rich trajectory,
$\alpha(M^2)$, whose real part is shown in Fig. \ref{fig:n1}.

Despite the seemingly linear form of the trajectory, it is not
that: the trajectory must contain an imaginary part corresponding
to the finite widths of the resonances on it. The non-trivial
problem of combining the nearly linear and real function with its
imaginary part was solved in Ref. \cite{Paccanoni} by means of
dispersion relations.

%%%%%%%%%%%%%%%%%%
%%%%%%%%%%%%%%%%%%

We use  the explicit form of the trajectory derived in Ref.
\cite{Paccanoni}, ensuring  correct behaviour of both its real and
imaginary parts. The imaginary part of the trajectory can be
written in the following way:
\begin{equation}
{\cal I}m\, \alpha(s)=s^{\delta} \sum_n c_n
\left(\frac{s-s_n}{s}\right)^{\lambda_n} \cdot \theta(s-s_n)\,,
\label{b2}
\end{equation}
where $\lambda_n={\cal R}e\ \alpha(s_n)$. Eq.~(\ref{b2}) has the
correct threshold behaviour, while analyticity requires that
$\delta <1$. The boundedness of $\alpha(s)$ for $s \to \infty$
follows from the condition that the amplitude, in the Regge form,
should have no essential singularity at infinity in the cut plane.

The real part of the proton trajectory is given by
\begin{equation}
{\cal R}e\,\alpha(s)=\alpha(0)+\frac{s}{\pi}\sum_n c_n {\cal
A}_n(s)~, \label{b3}
\end{equation}
where
\begin{eqnarray*}
& & {\cal A}_n(s)=
\frac{\Gamma(1-\delta)\Gamma(\lambda_n+1)}{\Gamma(\lambda_n-\delta+2)
s_n^{1-\delta}}{}_2F_1\left(1,1-\delta;\lambda_n-\delta+2;\frac{s}{s_n}\right)\theta(s_n-s)+\\
& & \left\{ \pi s^{\delta-1}\left(
\frac{s-s_n}{s}\right)^{\lambda_n} \cot[\pi(1-\delta)]- \right. \\
& & \left.
\frac{\Gamma(-\delta)\Gamma(\lambda_n+1)s_n^{\delta}}{s\Gamma(\lambda_n-\delta+1)
}{}_2F_1\left(\delta-\lambda_n,1;\delta+1;\frac{s_n}{s}\right)
\vphantom{\left( \frac{s-s_n}{s}\right)^{\lambda_n}} \right \}
\theta(s-s_n)~.
\end{eqnarray*}

As already mentioned, the proton trajectory, also called $N^+$
trajectory \cite{Paccanoni}, contains the baryons N(939)
$\frac{1}{2}^+$, N(1680) $\frac{5}{2}^+$, N(2220) $\frac{9}{2}^+$
and N(2700) $\frac{13}{2}^+$ \cite{particles}. In the fit, the
input data are the masses and widths of the resonances. The
quantities to be determined are the parameters $c_n$, $\delta$ and
the thresholds $s_n$. Following \cite{Paccanoni} we set $n=1,2,x$
and $s_1=(m_{\pi}+m_N)^2=1.16$ GeV$^2$, $s_2$ = 2.44 GeV$^2$ and
$s_x$ = 11.7 GeV$^2$.

Other parameters of the trajectory, obtained in the fit, are
summarized below: $\alpha(0)=-0.41$, $\delta = -0.46 \pm 0.07$,
$c_1=0.51\pm 0.08$, $c_2=4.0\pm 0.8$ and $c_x=(4.6 \pm 1.7)\cdot
10^{3}$. Taking the central values of these parameters we obtain
the following values for the $\lambda$'s: $\lambda_1=0.846$,
$\lambda_2=2.082$,  $\lambda_x=11.177$.

The fit is fairly good: $\chi^2/$d.o.f = 1.15, see
Figs.~\ref{fig:n1} and \ref{fig:n1width}. In the mass range were
the parameters of the trajectory were fitted to the data, i.e.
$M_X^2\le 8$ GeV$^2$, this is the most realistic proton trajectory
we know from the literature. Nevertheless, care should be taken if
used outside this range. As long as we are within our
applicability range, the sum over resonances in Eq. (\ref{Magas})
is restricted to 4 resonances ($n=0,3$), but in the imaginary part
of the transition amplitude, Eq. (\ref{ImA}) we consider the
contributions only from three of these resonances, since for the
lowest resonance, i.e. for the proton, $n=0$, the imaginary part
vanishes, $Im\, \alpha=0$, producing an infinitely narrow and high
peak.

\begin{figure}[htb]
\includegraphics[width=.6\textwidth]{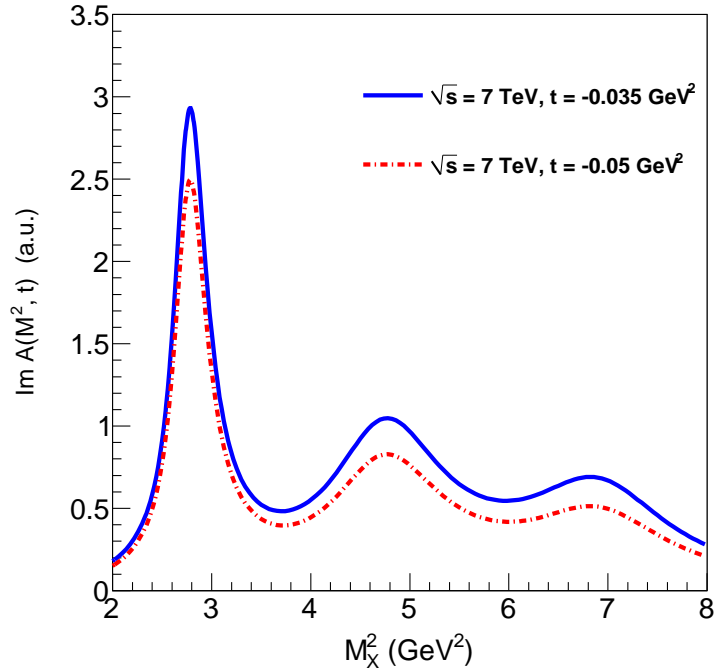}
\caption{Imaginary part of the amplitude $A(M^2_X,t)$, eq.
(\ref{ImA_fin}).} \label{fig:cross}
\end{figure}

The elastic contribution, $pP\rightarrow pP$ will be discussed in
the next section,  see also \cite{Bodek, Dissertation}. However,
it can be assumed that outside the elastic peak, $2$ GeV$^2$ $\le M_X^2\le 8$ GeV$^2$,
this distribution can be neglected, because the dominant part come from the nearest
resonance.

\begin{figure}[htb]
\includegraphics[width=.75\textwidth]{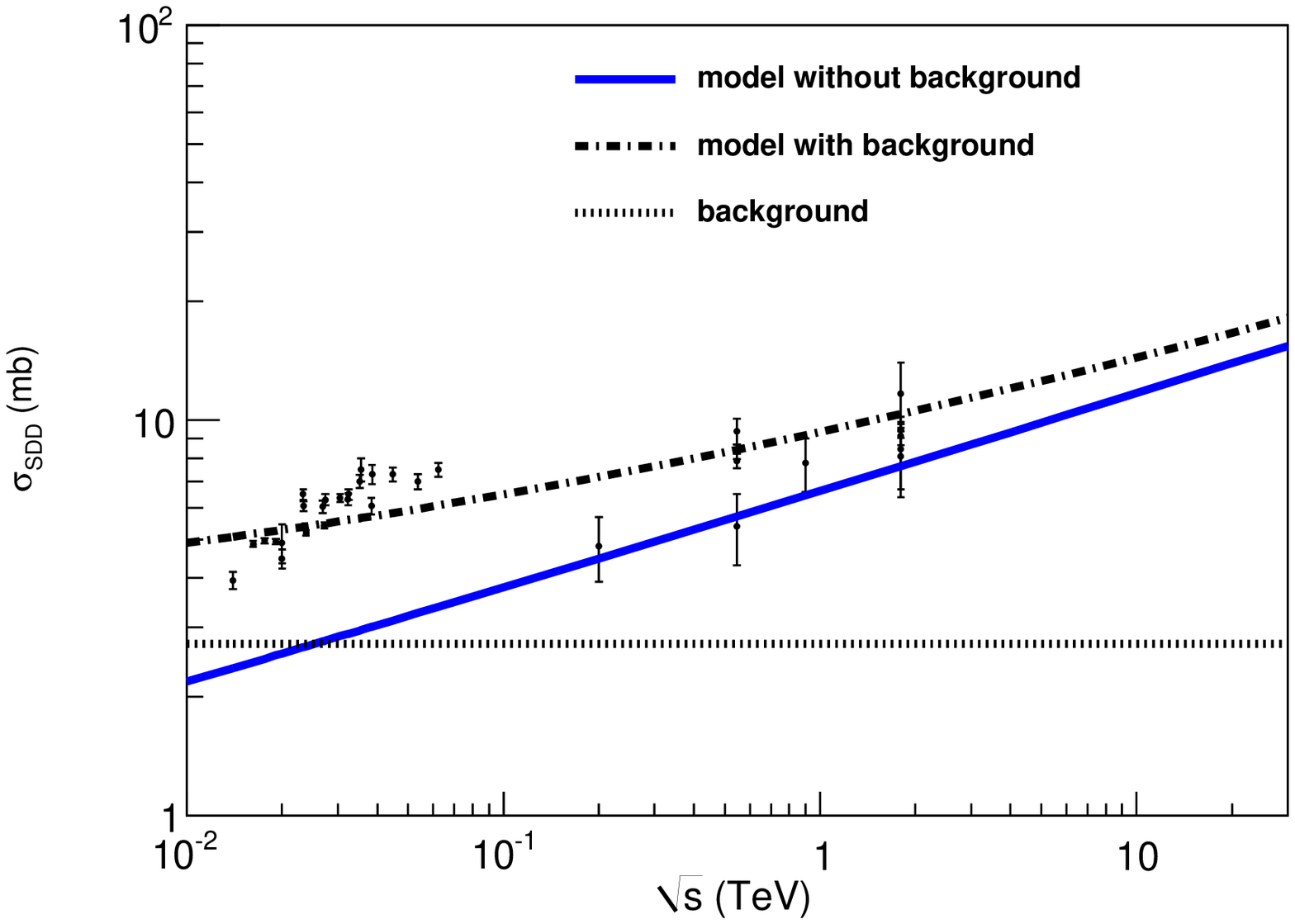}
\caption{Predicted integrated SDD cross section as a
function of $s$ compared with the experimental data \cite{Scham1,
Scham2, Albrow, Ansorge, Alner, Abe, Cool, Bernard, Amos_E710}; see also \cite{Arkhipov}.
\label{fig:2}
}
\end{figure}

Thus, we obtain:
\begin{equation}
\label{ImA_fin}
Im\,  A(M_X^2,t)=a \sum_{n=1,3}[f(t)]^{2(n+1)}
\frac{Im\, \alpha(M^2_X)}{(2n+0.5-Re\, \alpha(M_X^2))^2+
(Im\, \alpha(M_X^2))^2}\,.
\end{equation}
Note that the contribution from each subsequent resonance of the
proton trajectory is suppressed by a factor $f(t)^2$ compared with
the previous one.

Apart from the well established proton trajectory, with a sequence
of four particles on it, there is a prominent resonance $I=1/2,\ \
J=1/2^+$ with mass $1440$ MeV, known as the Roper resonance. It is
wide, the width being nearly one quarter of its mass. The Roper
resonance may appear on the daughter trajectory of $N^*$ treated
above, although its status is still disputable. In the Appendix
we consider the possible contribution from the single
Roper resonance by means of a separate Breit-Wigner term.

\begin{figure}[htb]
\includegraphics[width=.49\textwidth]{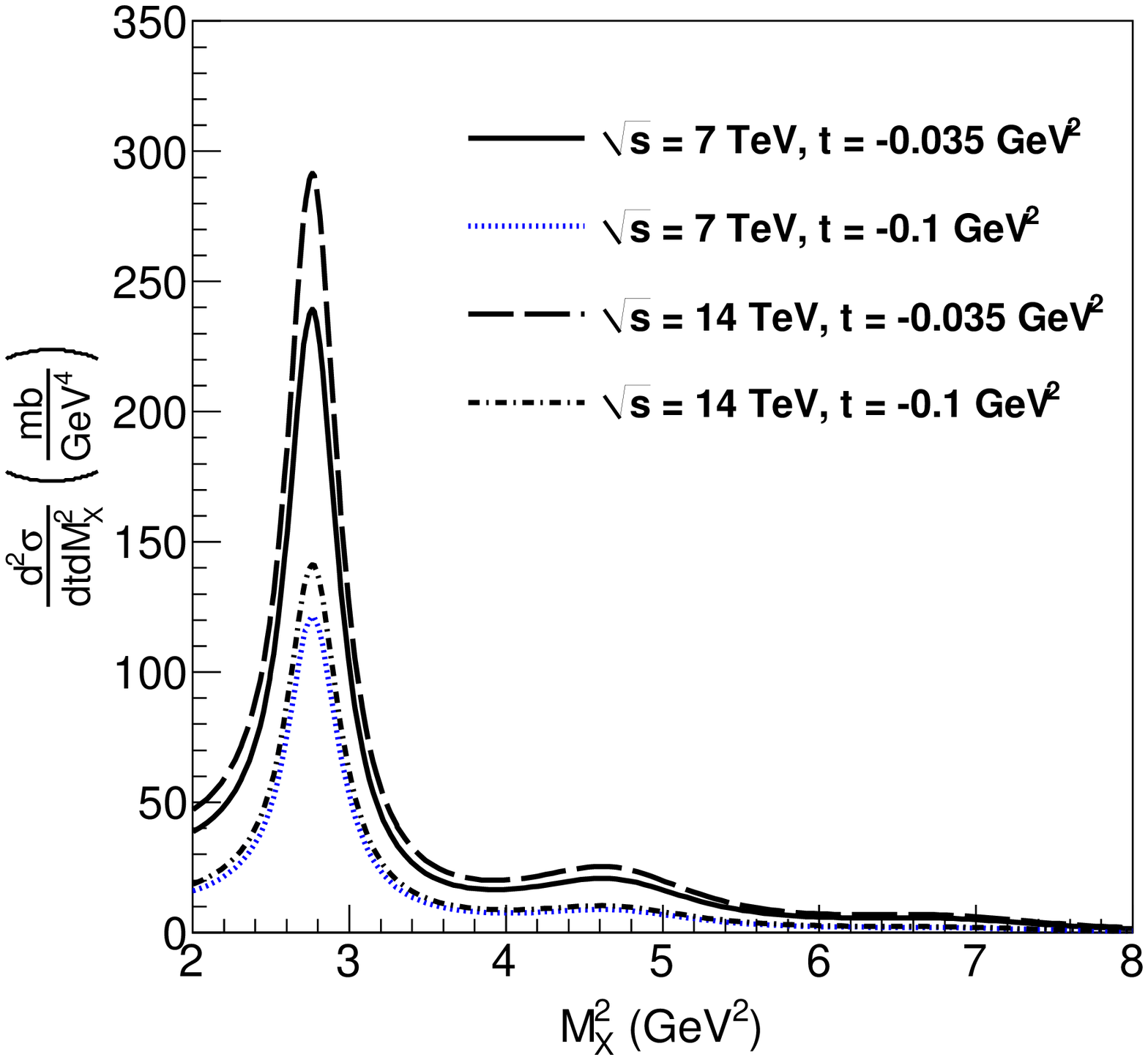}
\includegraphics[width=.49\textwidth]{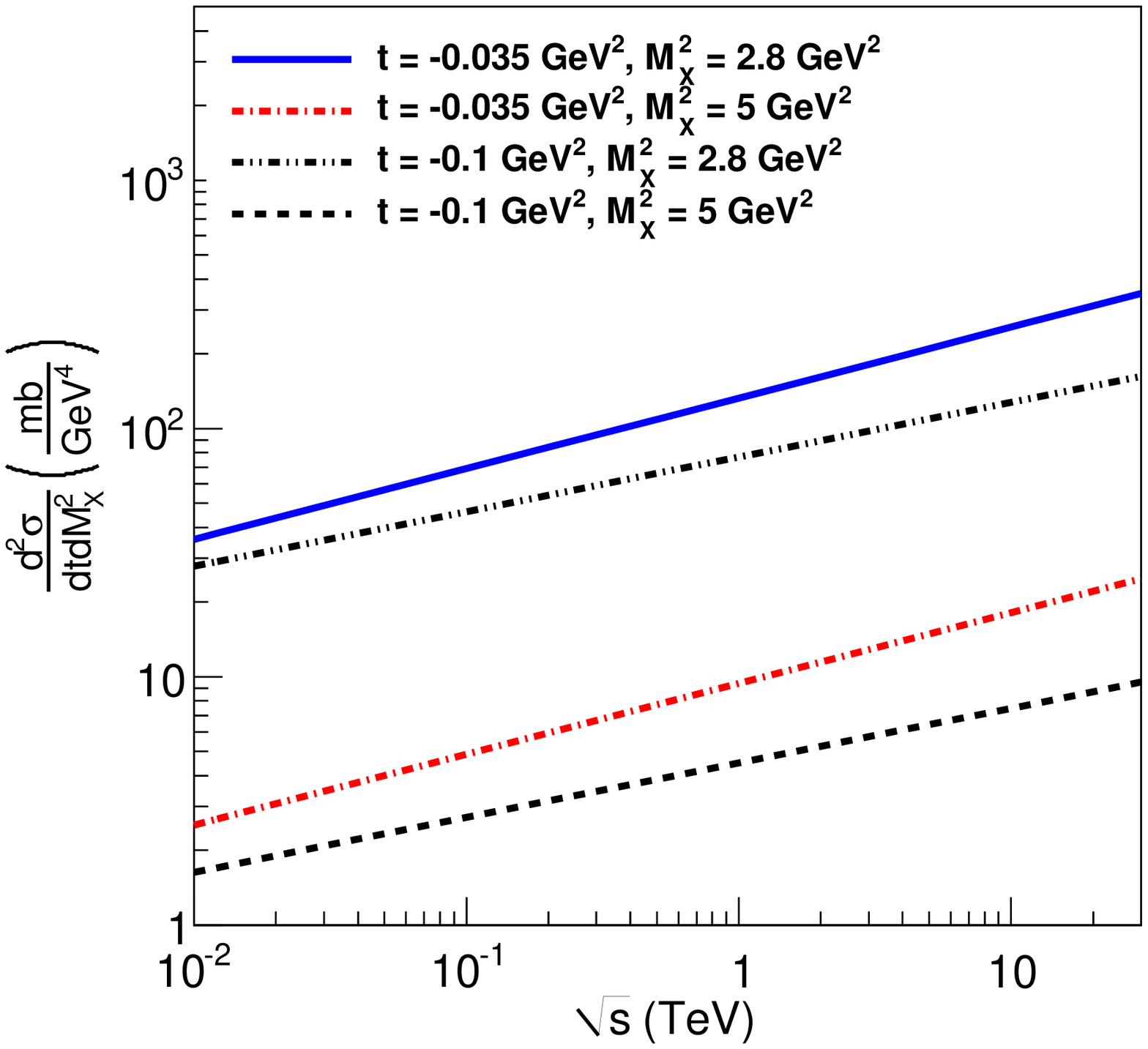}
\parbox[t]{8.5cm}{\caption{Double differential cross section of SDD as a
function of $M_X^2$ for several fixed values of $t$ and two
representative LHC energies.  \label{fig:3}
}}
\hfill~\parbox[t]{8.5cm}{\caption{Double-differential cross section of SDD as a
function of $s$ at fixed values of $t$ and $M^2_X$.
\label{fig:5}
}}
\end{figure}

\section{Results}\label{Results}

Fig. \ref{fig:cross} shows the behavior of the imaginary part of
the transition amplitude, eq. (\ref{ImA_fin}), proportional to the transition form
factor as in Fig. \ref{fig:drow1}, or lower vertex in Fig.
\ref{fig:diagram} (right panel). It shows the resonance structure
corresponding to the proton trajectory, to be translated into the
cross sections via eq. (\ref{DD2},\ref{eq11}), with the results shown below, eq. (\ref{DD2_fin}).
One can see that the imaginary part of transition amplitude
decreases with growing $|t|$, due to the dipole form factor
$f(t)^{2(n+1)}$. Furthermore
for each fixed $t$ the relative contribution of higher resonances
decreases, because of the suppression factor
 $f(t)^2$ for every subsequent resonance. Such a behaviour results
 from dual models.

\begin{figure}[htb]
\includegraphics[width=.49\textwidth]{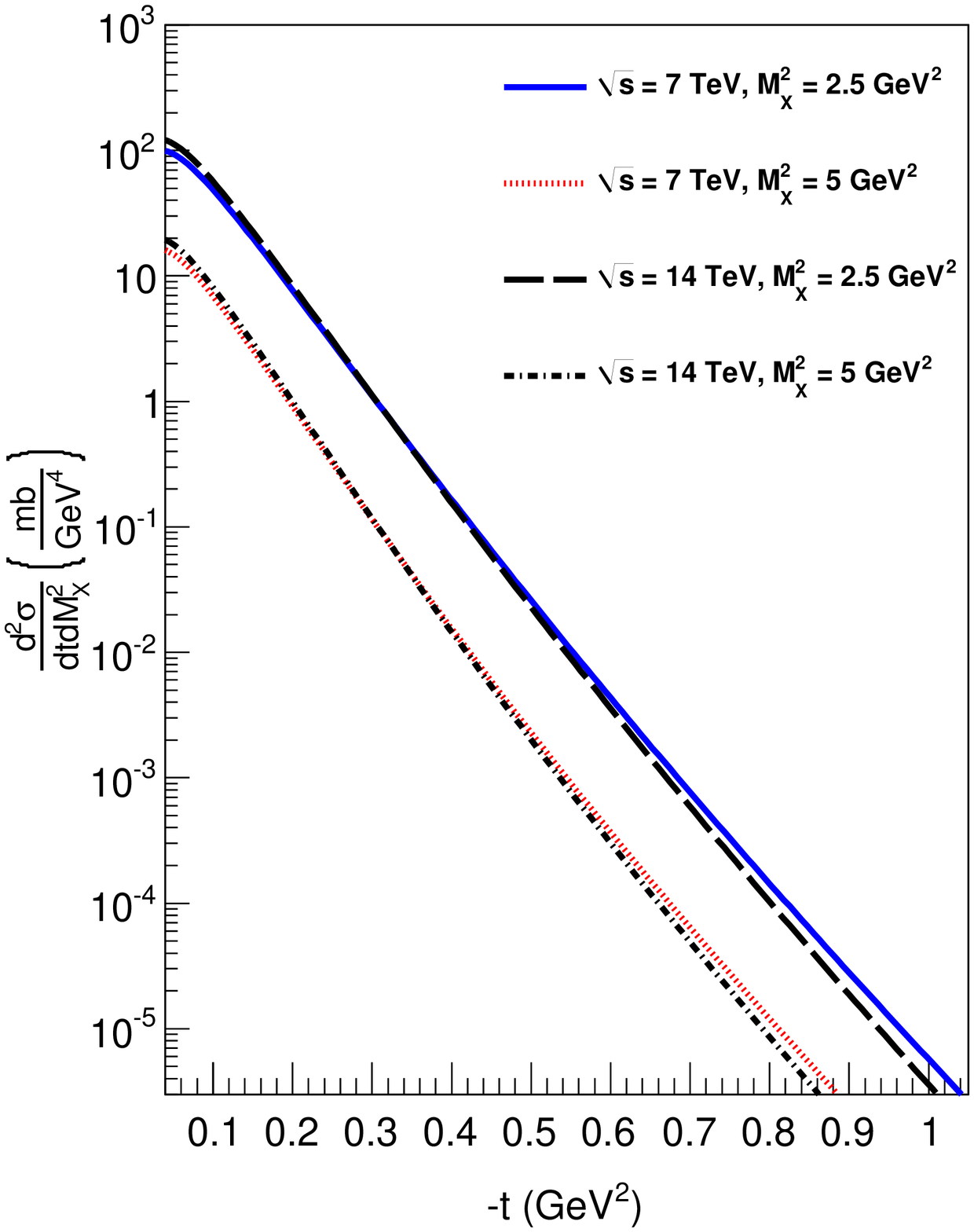}
\includegraphics[width=.49\textwidth]{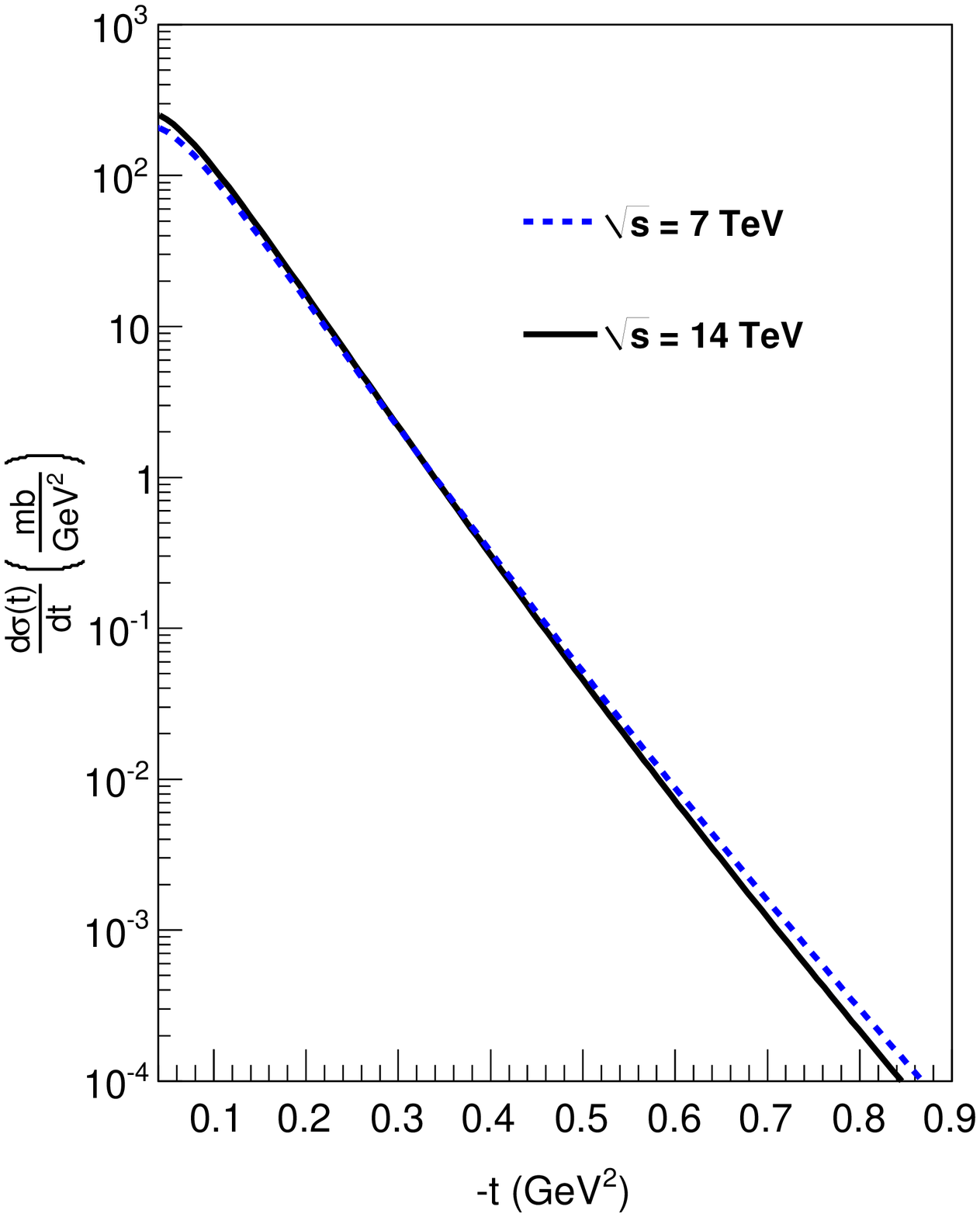}
\parbox[t]{8.5cm}{\caption{Double-differential cross section of SDD as a
function of $t$ at fixed values of $s$ and $M^2_X$.  \label{fig:4}}}
\hfill~\parbox[t]{8.5cm}{\caption{Integrated cross section
$d\sigma/dt$ as a function of $t$. \label{fig:4_1}}}
\end{figure}

Our final expression for the double differencial cross section reads:
\begin{equation}
\label{DD2_fin}
 \frac{d^2\sigma}{dtdM_X^2}=
A_0\left(\frac{s}{M_X^2}\right)^{2\alpha_P(t)-2}
\frac{x(1-x)^2\ [F^p(t)]^2}{(M_x^2-m^2) \left(1+\frac{4m^2x^2}{-t}\right)^{3/2}}
 \sum_{n=1,3}
\frac{[f(t)]^{2(n+1)}\, Im\, \alpha(M^2_X)}{(2n+0.5-Re\, \alpha(M_X^2))^2+
(Im\, \alpha(M_X^2))^2}\,.
\end{equation}
Its overall normalization depends on two factors, namely $\beta$, Eq. (\ref{DD2}), and $a$,  Eq. (\ref{Magas}),  but, since they do not appear separately, we have combined them in a single factor,
$A_0=\frac{9 a \beta^4}{\pi \alpha}$, to be fitted to the data. 

Data on the integrated SDD cross sections for different $s$ are
available from \cite{Scham1, Scham2, Albrow} (see also
\cite{Arkhipov}). Unfortunately, these data
points are not sufficient to fix this  norm unambiguously and to discriminate
uniquely the resonance contribution from the background.
Furthermore, in the present
model, applicable at the LHC, only the Pomeron  trajectory
is considered in the $t$ channel. At much lower energies, where data are available: Fermilab,
ISR,... -  secondary, non-leading trajectories give some contribution as well. 
They should be included in a future, more refined, analysis (fit) of these data. 

To calculate the integrated SDD we first take
into account the contribution from the resonance region. This is
done by integrating eq. (\ref{DD2_fin}) in squared momentum transfer
$t$ from $-s$ to $0$, and in the missing mass $M_x$ over the
resonance region, $2$ GeV$^2$ $<M_x^2<8$ GeV$^2$, where the
contributions from the resonances, eq. (\ref{DD2}) dominate.
We, thus, eliminate contributions from the region of the elastic peak, $M_X^2<2$ GeV$^2$, that
requires separate treatment, see \cite{Otranto}, and the high missing mass
Regge-behaved region. By duality, to avoid "double counting", the latter should be
accounted for automatically, provided the resonance contribution is included properly.

The results for the integrated SDD cross section are shown in Fig.
\ref{fig:2}. Without any background contribution, fits to the data
give $A_0= 977 \pm 5$ mb/GeV$^2$ with $\chi^2/d.o.f.=25.7$. Better agreement with
the data can be obtain by including a constant background, i.e. by
adding a fitting parameter $b$ to the integrated SDD cross
section. In this case, the fit gives $\chi^2/d.o.f.=11.5$, with
$A_0=506 \pm 23$ mb/GeV$^2$ and $b=2.72 \pm 0.13$ mb. Lacking any reliable model for the 
background, we avoid complicated background parameterization. Our constant boackground gives about 20$\%$ contribution at LHC energy, that seems to be a reasonable number. 

Having fixed the parameters of the model, we can now scrutinize
the SDD cross section in more details. First we calculate double
differential cross section, eq. (\ref{DD2_fin}), as a function of the
missing mass for several fixed values of the momentum transfer $t$
and two representative LHC energies, $7$ and $14$ TeV. The results
of such calculations are shown in Fig. \ref{fig:3}.

In Fig. \ref{fig:5} we show the energy dependence of the double
differential cross section for several fixed values of $t$ and $M_X$. The rise of $\sigma_{SDD}(s)$ is mainly determined by the supercritical Pomeron intercept $\alpha_P(0)$, and only weekly affected also by the details of the $t-$ dependence of the Pomeron trajectory.

The double differential cross sections as a function of $t$ for two 
representative LHC energies and several fixed values of $M_X$ 
are presented in Fig. \ref{fig:4}.  
Fig. \ref{fig:4_1} shows the $t$- dependence of the differential cross
section integrated in $M_X^2$ for representative LHC energies. 

\section{Conclusions}

Let us briefly summarize the status of the present model and its
credibility, including the way its parameters were fixed. As
already mentioned, the normalization constant, $\beta$, discussed in
Secs. \ref{s2} and \ref{SDD} is absorbed by the overall norm $A_0$, together with
the other normalization paremeter $a$.  The parameters of the Pomeron
trajectory were determined \cite{L} from $pp$ elastic and total cross
section data. The form and the values of the parameters of the proton
trajectory, that plays a crucial role in predicting the $M_X$
dependence, are fixed by spectroscopic data, see Sec.
\ref{Sec:trajectory}. Finally the overall norm $A_0$ is fixed from
the comparison of calculated SDD cross section with the
experimental data, with the following caveat: in the present
model, applicable at the LHC, only the Pomeron  trajectory
contributes in the $t$ channel. At much lower energies (Fermilab,
ISR,...), where data are available, apart from the Pomeron,
secondary, non-leading trajectories contribute as well. We plan to included these 
in a future,  more refined, study. 

There is some freedom in the form and weight of the background.
Its relative contribution can only be normalized to earlier
measurements at the ISR or the Fermilab. For a better control, we compare our
predictions with the experimental data \cite{Scham1, Scham2,
Albrow} and theoretical estimates \cite{Goulianos, Montanha,
Arkhipov}. In any case, it follows from our model and the fits to the data
that the background is fairly large: about $20 \%$ at the LHC. A dedicated study
of various options for the background in SDD can be found in Ref. \cite{Otranto}.

The elastic contribution, $pp\rightarrow pp$ is usually calculated
and measured separately. There is no consistent theoretical
prescription of any smooth transition from inelastic to elastic
scattering, corresponding to the $x\rightarrow 1$ limit for the
structure functions (see Ref. \cite{Otranto}).

There is an important point omitted in this short paper, namely
unitarity. As is well known (see, e.g. \cite{Collins}) any simple
Regge pole model violates unitarity in the sense that the DD cross
section asymptotically grows faster that the total cross section
(it is obvious that no partial cross section can overshoot the
total cross section). This long-standing problem was cured in
various ways, the final answer being still open. In Refs.
\cite{Goulianos, Goulianos1, Montanha} unitarity is restored by a
renormalization procedure. Without entering into details, here we
only mention that a possible solution of this problem can be found
by using a more realistic (and complicated) Pomeron singularity,
for example in the form of a double pole \cite{DP}.

The model presented in this paper and the calculated cross
sections, corrected for the efficiencies of relevant detectors
will be used \cite{future} in future measurements at the LHC.

As already mentioned in the Introduction, the prospects of
measuring SDD at the LHC are promising, although some detail still
remain to be settled. For the CMS Collaboration, the SDD mass
coverage is presently limited to some 10 GeV. Together with the T2
detectors of TOTEM, SDD masses down to 4 GeV could be covered,
provided the TOTEM trigger (data acquisition) system will be
combined with the CMS ones. ALICE and LHCb have different beam
arrangements, but their acceptances for central diffraction
(double pomeron exchange) was also investigated (see, e.g.
\cite{Risto}). Measurements of the SDD events at the LHC are based
on: (1) identifying a gap in forward rapidities in conjunction
with a veto for any activity on the opposite side of the
interaction point, or (2) detecting a diffractively scattered
proton in a leading proton detector, such as the Roman Pots, and a
coincident diffractively excited bunch of particles on the
opposite side. The problem with both measurement strategies stems
from the incomplete rapidity coverage of the base line detector
systems at the LHC: the low mass, $M < 4$ GeV, diffractively excited
states are not seen. Without extra rapidity coverage below M = 4
GeV, both approaches to SDD identification fail. In case of purely
rapidity gap based method, the recorded cross section misses the
SDD events with diffractive masses below 4 GeV. In case a leading
proton is detected on one side of the Intersection Point (IP), one
could, in principle, be sensitive to diffractive masses that
correspond to the uncertainty in LHC beam energy. In practice, it
is impossible to trigger for these events, and the low mass SDD
events will be missed by this method as well. Detecting SDD events
with high acceptance is essential for determining the total pp
cross section in the so called luminosity independent method based
on using the Optical Theorem. The method bases on measuring the
slope of the elastic cross section, extrapolating the slope to the
optical point. Together with the over-all inelastic rate (plus the
ratio between the inelastic and elastic forward scattering
amplitudes), the total pp cross section is obtained. The main
uncertainty in this evaluation is due to the error in estimating
the inelastic pp event rate. As shown by the authors of Ref.
\cite{Risto}, the acceptance of basically all the LHC experiments
can be substantially improved by adding forward detector systems
(Forward Shower Counters, FSCs) that register secondary
interactions within the beam pipe due to particles - both
electrically neutral and charged - emitted at very small
scattering angles with respect to the beam direction. With the
addition of FSCs, rapidity coverage of an LHC experiment can be
extended down to SDD masses of the order of 1.2 GeV, i.e. down to
the dominant N* states. FSCs are being currently installed in
ALICE and CMS detectors, and they will provide the necessary added
coverage of small mass forward systems at the LHC.

\section*{Acknowledgements}
We thank K. Goulianos and F. Paccanoni for useful discussions.
R.O. would like to acknowledge his thanks
to the Academy of Finland for support. O.K. is grateful to Rainer Schicker and A.G. Zagorodny for their support; acknowledges the hospitality of the Bogolyubov Institute for Theoretical Physics (Kiev, Ukraine); his work was supported by the
EU program "Difractive and Electromagnetic Processes at the LHC".
The work of V.K.M. is partly supported by the contracts FIS2008-01661 from
MICINN (Spain), by the Ge\-ne\-ra\-li\-tat de Catalunya contract
2009SGR-1289, and by the European Community-Research
Infrastructure Integrating Activity ``Study of Strongly
Interacting Matter'' (HadronPhysics2, Grant Agreement n. 227431)
under the Seventh Framework Programme of EU. L.J. was supported by the National Ac. Sc.
of Ukraine, Dept. of Astronomy and Physics, under the Grant "Matter at Extreme Condicion".

\section{Appendix. Roper}
\label{AppB}

Apart from the well established protonic trajectory
with a sequence of four particles, on it Sec.
\ref{Sec:trajectory}, there is a prominent single resonance
$I=1/2,\ \  J=1/2^+$ with mass $1440$ MeV, known as the Roper
resonance \cite{particles}. It is wide, the width being nearly one
fourth of its mass, its spectroscopic status being disputable.
There is no room for the Roper resonance on the proton trajectory
of Sec. \ref{Sec:trajectory}, although it could still be a member of protons
daughter trajectory. Waiting for a future better understanding of Roper's status, here we
present the contribution to SDD cross section of a single Roper
resonance, calculated from a simple
Breit-Wigner formula:

\begin{figure}[h]
\includegraphics[width=.49\textwidth]{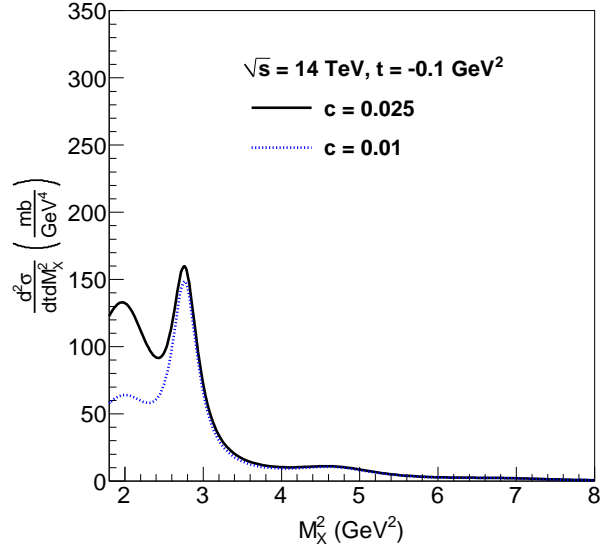}
\caption{Double-differential cross section of SDD as a function of
$M^2_X$ at fixed values of $s$ and $t$, with the Roper resonance
added according to eq. (\ref{Roper_eq}). \label{fig:6} }
\end{figure}

\begin{equation}
\label{Roper_eq}
Im\,  A_{incl. Roper}(M_X^2,t)= a \left( \sum_{n=1,3}
\frac{[f(t)]^{2(n+1)}\, Im\, \alpha(M^2_X)}{(2n+0.5-Re\, \alpha(M_X^2))^2+
(Im\, \alpha(M_X^2))^2}
+ c \frac{f^2(t) M_{Roper} \Gamma_{Roper}/2 }{(M^2_X - M^2_{Roper})^2 +
(\Gamma_{Roper}/2)^2} \right) \,,
\end{equation}
where $M_{Roper} = 1440$ MeV, $\Gamma_{Roper} = 325$ MeV,  and $c$ is another normalization parameter. For illustration we will take two different values of $c$; the resulting shapes of the double-differential cross section are presented at Fig. \ref{fig:6}.

\vfill \eject
\end{document}